\documentstyle[12pt]{article}
\def\be{\begin{equation}}
\def\ee{\end{equation}}
\def\beqn{\begin{eqnarray}}
\def\eeqn{\end{eqnarray}}
\def\noin{\noindent}
\begin{document}
\begin{titlepage}
\title{Loop Equations for the d-dimensional\\ One-Hermitian Matrix model}
\author{Jorge Alfaro\thanks{Permanent address: Fac. de F\'\i sica, Universidad
Cat\'olica de Chile, Casilla 306, Santiago 22, Chile.}
\\Theory Division, CERN\\CH-1211 Geneva 23\\
and\\
Facultad de F\'\i sica\\Universidad Cat\'olica de
Chile\\Casilla 306, Santiago 22, Chile
}

\maketitle
\begin{abstract}

We derive the loop equations for the one Hermitian matrix model in any
 dimension. These  are a consequence of the Schwinger-Dyson equations of the
 model. Moreover we show that in leading order of large $N$ the loop equations
 form a closed set.
\end{abstract}
\vfill
\begin{flushleft}
CERN--TH-6966/93 \\
August 1993
\end{flushleft}
\end{titlepage}
\newpage

\section{Introduction}

Recently enormous progress  has been made in the
understanding of the physics of non-perturbative strings
in a low number of dimensions. This has been made possible by the
matrix model technology. The Feynman graphs
of the matrix models provide a suitable discretization
of the two-dimensional surface where the string lives,
classified according to its genus, which in the matrix model is a
given order in $1/N$ ($N$ being the range of the matrix). By tuning
the couplings in the matrix model in such a way that the perturbative
series diverges, the continuum limit of the two-dimensional
surface can be approached. In this way
it is possible to get a sum over all topologies  in  the  first  quantized
string,  using  as
a   discretization   of  the  two-dimensional  world-sheet  a
dense  sum  of Feynman  graphs  of  an  associated  matrix  model.

Whenever the matrix model is solvable, the corresponding
string properties can be also determined. In this way $2d$
gravity coupled to conformal matter with $c\le 1$  has been analysed and
the critical exponents determined \cite{volodya}\cite{gross} \cite{moore}.

Furthermore it has been shown that a Virasoro algebra naturally
arises in these matrix models \cite{verlinde}. This algebra is a realization
of the
Schwinger-Dyson equations in the $U(N)$ invariant set of operators of
the matrix model. Using this algebra, the equivalence of zero-dimensional
matrix models to topological
$2d$ gravity has been proved \cite{witten2}. Some time ago, an application of
these
ideas to compute intersection indices on moduli spaces of Riemann surfaces
has been
made \cite{kontsevich}.

 More recently further progress has been made possible by the
complete solution of the two-matrix model in zero
 dimensions\cite{metha}\cite{alf}\cite{staudacher}.

To study non-critical strings in higher dimensions, a knowledge of the solution
 of
matrix models in $d>1$ is needed, $d$ is the dimension of space-time,\ (for a
 different approach
to the problem see \cite{hikami}). But beyond $d=1$ the matrix
model is not solvable (although the gauged matrix model is
\cite {kazakov}). So various approximation schemes have been
suggested to tackle this problem \cite{luis}.

In this note we want to derive the loop equations for
the Hermitian matrix model for arbitrary $d$.
They are a consequence of the Schwinger-Dyson equations
that its correlation functions satisfy. Moreover we will
show that in the large $N$- limit, the loop equations form a
closed set. We believe this is an important step in the
understanding of the physics of these models because
recent experience has shown \cite{verlinde} that the
Schwinger-Dyson equations are a powerful tool to explore
the involved symmetries . It would be desirable to explore
the loop equations we present here from this perspective.

A simple and powerful way to derive Schwinger-Dyson equations
is the hidden BRST method \cite{aldamg}. In fact this
 was used to find the loop equa\-tions for the zero-dimensional, two-matrix
 model in  \cite{alf}.
So we start by reviewing this method  in section 2 and apply it to
derive the loop equations of the $d$-dimensional matrix model
in the next section. In section 4 we consider the case $d=0$
as a check. In section 5 we draw some conclusions and
comment on future work.

\section{Hidden BRST Symmetry and Schwinger-Dyson Equations}

The common procedure to derive Schwin\-ger-Dyson equations (SDe) involves
using the invariance of
the path integral measure under field translations.  Invariances of
the action itself are not relevant for this derivation.  In this
section we review a different approach which has been
proposed some years ago \cite{aldamg}: By making use of a (trivial) BRST
 symmetric extension of
any action $S$, we can derive all SDe's as BRST supersymmetric Ward
identities of the new action.
\vskip 1pc
For the sake of simplicity consider the path integral describing
one bosonic field
$\varphi(x)$:

\be
Z = \int[d\varphi] e^{-S(\varphi)} .
\ee
\noin
The functional measure $[d\varphi]$ in the last equation is invariant
under field translations:
\be
\varphi(x)\rightarrow\varphi(x) + \epsilon(x) .
\ee
\noin
The invariance of the functional measure implies the following
statement:
\be
\int[d\varphi]\frac{\delta[Fe^{-S(\varphi)}]}{\delta\varphi(x)} = 0
\ee
\noin
{}From (3) we get
\be
\langle\frac{\delta F}{\delta\varphi(x)}-F\frac{\delta
 S}{\delta\varphi(x)}\rangle = 0
\ee
\noin
for any $F=F(\varphi)$.  These are the SDe's
for the theory.  Notice
that, in general, $S$ is not invariant under (2) and that this lack
of invariance plays no role in the derivation of the SDe's  We now
introduce auxiliary fermionic variables $\psi(x)$, $\bar \psi(x)$ and insert
a trivial factor of unity
$$
1 = \int [d\bar\psi][d\psi] e^{-\int dx\bar\psi(x)\psi(x)}
$$
\noindent
inside the partition function.  The resulting action $\bar S$ is
invariant under the following transformation:
 \beqn
\bar S [\varphi,\psi,\bar\psi] &= S[\varphi] + \int dx \bar
\psi(x)\psi(x)\\
\delta \varphi(x) &= \psi (x) \\
\delta \psi (x) &= 0\\
\delta \bar \psi(x) &= -\delta S/\delta\varphi(x)
\eeqn

\noin
Associated with the BRST-like symmetry, (6)--(8) is a set of Ward
identities; for example, unbroken BRST implies:

$$
\langle\delta [F(\varphi) \bar\psi(y)]\rangle = 0,
$$
\noin
which is
$$
\langle\int dx \frac{\delta F}{\delta \varphi(x)}\psi(x)
    \bar\psi(y)
    -F \frac{\delta S}{\delta \varphi} \rangle=0 .
$$

\noin
Computing the average with respect to the fermionic variables we recover
eq. (4).  Notice that the BRST transformation we have just
defined commutes with any symmetry the original action $S$ may have.
In particular, if the original action is invariant under a $U(N)$ group, the
 BRST
symmetry will commute with this group also and then its implications will be
true order by order in the $1/N$ expansion.

In the next section we use  the Schwinger-Dyson BRST  symmetry  to
derive the loop equations for the $d$-dimensional Hermitian matrix
model.

\section{Loop Equations}

In this section we will derive the loop equations for the model defined
by the action:
\be
S=\int d^dx tr\left(\frac{1}{2}\partial_\mu M(x)\partial_\mu
M(x)+V(M(x))\right)
\ee
$M(x)$ is an $N\times N$ Hermitian matrix defined on the Euclidean space-time
 point
$x$, and $V$ is a local function of $M$ (normally a polynomial whose
degree can be
restricted by renormalizability of the model for $d>2$).

The extended action and the BRST transformation are:
\beqn
S_{ext}&=S+\int d^dx tr\bar\psi(x)\psi(x)\\
\delta M(x)&=\psi(x)\\
\delta\psi(x)&=0\\
\delta\bar\psi(x)&=-\frac{\delta S}{\delta M(x)}
\eeqn

Introduce the following representation of $M(x)$:
\be
M(x)=\sum_{\alpha=1}^N m_\alpha(x)T^\alpha(x),
\ee
where the projectors of the matrix $M(x)$ satisfy:
\beqn
\sum_{\alpha=1}^N T^\alpha(x)&=1\\
tr T^\alpha(x) T^\beta(x)&=\delta_{\alpha\beta}\\
T^\alpha(x) T^\beta(x)&=\delta_{\alpha\beta} T^\alpha(x) .
\eeqn
If $\delta$ denotes the BRST variation (see last section), we get the following
BRST transformation for the projectors $T^\alpha$ and eigenvalues $m_\alpha$
(see the Appendix):
\beqn
\delta T^\alpha(x) &= \sum_{\beta \neq \alpha}\frac{T^{\alpha}(x) \psi(x)
T^{\beta}(x) + T^{\beta}(x)\psi(x) T^{\alpha}(x)}{m_{\alpha}(x)-m_{\beta}(x)}
\\
\delta m_\alpha (x) &= {\rm tr}\psi(x) T^\alpha(x) .
\eeqn

Consider the following functional of $\lambda(x)$
\be
K_{ij}=\int dx \sum_{\alpha=1}^N\lambda_\alpha(x) T^\alpha(x)_{ij} .
\ee
Notice that $K_{ij}$ is an $x$-independent $N\times N$
matrix.

The basic object we will use to write the loop equations is:
\be
u[\lambda]=e^{iK}
\ee
It fullfils the identity:
\be
\frac{\delta u[\lambda]}{\delta\lambda_\alpha(x)}=
\int_0^1 dt e^{itK} iT^\alpha(x) e^{i(1-t)K} ,
\ee
which implies:
\be
\frac{\delta {\rm tr} u[\lambda]}{\delta\lambda_\alpha(x)}=
i tr T^\alpha(x) u[\lambda]
\ee
The trace is taken with respect to the internal indices only.
{}From $u[\lambda]$ we can compute all symmetric combinations of products
of $T^\alpha$. For instance:
\be
\frac{\delta^2u[\lambda]}{\delta\lambda_{\beta_1}(y_1)
\delta\lambda_{\beta_2}(y_2)}|_{\lambda=0}=-\frac{
T^{\beta_2}(y_2)T^{\beta_1}(y_1)+T^{\beta_1}(y_1)T^{\beta_2}(y_2)}
{2}
\ee
The Schwinger-Dyson equation is:
\be
\langle\delta tr [u[\lambda]\bar\psi(y)]\rangle=0
\ee
i.e.
\be
\langle tr [\delta u[\lambda]\bar\psi(y)+tr [
 u[\lambda]\delta\bar\psi(y)]\rangle=0 .
\ee
Since
\be
\delta\bar\psi(y)=\Box_y M-\Delta(M)
\ee
where
\be
\Delta(M)=V'(M) .
\ee

We get:
\be
tr u [\Box_y M(y)-\Delta(y)]=\sum_\alpha[m_\alpha(y)\Box_y+2\partial_\mu
 m_\alpha(y)\partial_\mu+\Box_y m_\alpha(y)-\Delta_\alpha]
tr u T^\alpha(y) .
\ee
We can evaluate:
\be
tr u T^\alpha(y)=-i \frac{\delta tr u[\lambda]}{\delta\lambda_\alpha(y)} .
\ee

We can also compute $\delta u[\lambda]$:
\beqn
\delta u =\int^1_0 dt e^{i t \int dx \sum_{\alpha=1}^N\lambda_\alpha(x)
 T^\alpha(x)}
i \int dx \sum_{\alpha=1}^N \lambda_\alpha(x) \sum_{\beta\ne \alpha}\nonumber\\
\frac{T^\alpha(x)\psi(x)T^\beta(x)+
T^\beta(x)\psi(x)
T^\alpha(x)}{m_\alpha-m_\beta}
e^{i(1-t)\int dx \sum_{\alpha=1}^N \lambda_\alpha(x) T^\alpha(x)}
\eeqn

After computing the fermionic average, we get:
\be
\langle\delta u\bar\psi(y)\rangle= \langle i\sum_{\beta\ne\alpha}\int^1_0
 \frac{\lambda_\alpha(y)-
\lambda_\beta(y)}{m_\alpha-m_\beta} tr u[t\lambda] T^\beta(y)
tr u[(1-t)\lambda] T^\alpha(y)\rangle .
\ee

Since $T^\alpha(y)$ appears only once in the trace, we can express
it as a derivative of $u$:
\beqn
tr u[t\lambda] T^\beta(y) =\frac{-i}{t}\frac{\delta tr u[t\lambda]}
{\delta\lambda_\beta(y)} \nonumber\\
tr u[(1-t)\lambda] T^\alpha(y) =\frac{-i}{1-t}
\frac{\delta tr u[(1-t)\lambda]}{\delta\lambda_\alpha(y)}
\eeqn
Therefore the loop equation is:
\beqn
\langle\sum_{\beta\ne\alpha}\int_0^1 dt \frac{\lambda_\alpha(y)-
\lambda_\beta(y)}{m_\alpha-m_\beta}\frac{1}{t(1-t)}
\frac{\delta tr u[t\lambda]}{\delta\lambda_\beta(y)} \frac{\delta
tr u[(1-t)\lambda]}{\delta\lambda_\alpha(y)}\rangle= \nonumber\\
-\langle\sum_\alpha[m_\alpha(y)\Box_y+2\partial_\mu
 m_\alpha(y)\partial_\mu+\Box_y m_\alpha(y)-\Delta_\alpha]
\frac{\delta tr u[\lambda]}{\delta\lambda_\alpha(y)}\rangle ,
\eeqn
which is valid to any order in the $1/N$ expansion.

A simplification is possible if we restrict ourselves to the leading order
in $1/N$. In this case we can apply Witten's factorization
property  \cite{witten}to the
loop equations to prove that:
$$
F(M)\sim \langle F(M)\rangle + O(N^{-a}) ,\ \ a>0
$$
for any $U(N)$-invariant $F$.
In particular, we obtain that:
$$
m_\alpha(x)\sim\langle m_\alpha(x)\rangle=\langle m_\alpha(0)\rangle .
$$
Therefore, in leading order of large $N$ the loop equation becomes a closed
 system for $u[\lambda]$:
\beqn
\sum_{\beta\ne\alpha}\int_0^1 dt \frac{\lambda_\alpha(y)-
\lambda_\beta(y)}{m_\alpha-m_\beta}\frac{1}{t(1-t)}
\frac{\delta tr u[t\lambda]}{\delta\lambda_\beta(y)} \frac{\delta
tr u[(1-t)\lambda]}{\delta\lambda_\alpha(y)}\nonumber\\
=-\sum_\alpha(m_\alpha\Box_y-
\Delta_\alpha))\frac{\delta tr u[\lambda]}{\delta\lambda_\alpha(y)}
\eeqn

The reason why we could derive this closed system was that we chose to write
the Schwinger-Dyson (loop) equations in terms of $T^\alpha$ rather than $M$.
Notice also that by using  the BRST Schwinger-Dyson transformation we can avoid
 this difficult change of variables in the path integral and the (highly)
 non-trivial measure in the $(m_\alpha,T^\alpha)$ basis.
\vskip 1pc
The loop equations must be solved subject to the following
boundary conditions:
\beqn
tr u|_{\lambda=0}=N\nonumber\\
\frac{\delta tr u}{\delta\lambda_\alpha(x)}=i \nonumber\\
\frac{\delta^2 tr u}
{\delta\lambda_\alpha(x)\lambda_\beta(x)}|_{\lambda=0}=
-\delta_{\alpha\beta}
\eeqn

A particular solution of the loop equations is $u=$ constant ($\lambda$
independent), which does not satisfy the boundary conditions.
We must look for a non-trivial solution.

An  equivalent  formulation of the loop equations is  obtained  by
noticing  that  only  the derivative of $tr u$  appears  in  the  loop
equations. So identify:
\be
v[\lambda]_\alpha(x)=\frac{\delta tr u}{\delta\lambda_\alpha(x)} .
\ee
Then  the loop equation is equivalent to the following  system  of
equations:
\beqn
\sum_{\beta\ne\alpha}\int_0^1 dt \frac{\lambda_\alpha(y)-
\lambda_\beta(y)}{m_\alpha-m_\beta}\frac{1}{t(1-t)}
v[t\lambda]_\beta(y) v[(1-t)\lambda]_\alpha(y) \nonumber\\
=-\sum_\alpha(m_\alpha\Box_y-
\Delta_\alpha))v[\lambda]_\alpha(y)\\
\frac{\delta v[\lambda]_\alpha(x)}{\delta \lambda_\beta(y)}=
\frac{\delta v[\lambda]_\beta(y)}{\delta \lambda_\alpha(x)}\\
v[\lambda]_\alpha(x)|_{\lambda=0}=i\\
\frac{\delta
v[\lambda]_\alpha(x)}{\delta\lambda_\beta(x)}|_{\lambda=0}=
-\delta_{\alpha\beta}
\eeqn

Notice that:
\beqn
tr M(x_1) M(x_2) ...M(x_n)=\sum_{\alpha_1...\alpha_n} m_{\alpha_1}
...m_{\alpha_n} tr T^{\alpha_1}(x_1)...T^{\alpha_n}(x_n) \nonumber\\
=\frac{1}{N}\sum_{\alpha_1...\alpha_n} m_{\alpha_1}...m_{\alpha_n}
[tr T^{\alpha_1}(x_1) ..T^{\alpha_n}(x_n)+ \rm{  all\  permutations\  of
\ indices\  \alpha}]
\eeqn

Therefore it is enough to know $tr u[\lambda]$ to get all correlations of the
 matrix $M$ that are symmetric under permutations of the space-time points
 $x_i$.

We can get a continuum version of the large-$N$ loop equation by the usual
 manipulations. That is we introduce the functions $\lambda(x,y)$ and $m(x)$
 defined for $0\leq x\leq 1$ and the density of eigenvalues $\rho$, and make
the
 following identifications:
\beqn
\lambda_\alpha(y)=\lambda(\frac{\alpha}{N},y)\\
m_\alpha=\sqrt{N} m(\frac{\alpha}{N})\\
\rho(m)=\frac{dx}{dm}\\
\int\rho(m)=1 .
\eeqn

Then the large-$N$ loop equation becomes:
\beqn
P\int dm_1 dm_2\rho(m_1)\rho(m_2)\int_0^1
 dt\frac{\lambda(m_1,y)-\lambda(m_2,y)}{m_1-m_2}\frac{1}{t(1-t)}\nonumber
\\
\times\frac{\delta tr u[t\lambda]}{\delta\lambda(m_2,y)}\frac{\delta tr
 u[(1-t)\lambda]}{\delta\lambda(m_1,y)}=-\int dm(m\Box_y-\Delta(m))
\frac{tr u[\lambda]}{\delta\lambda(m,y)} ;
\eeqn
$P$ stands for the principal value of the integral over $m_i$.

In the next section we will solve in detail the $d=0$ case and show that it
 agrees with the known solution \cite{brezin}.

\section{Zero-dimensional matrix model}

The zero-dimensional matrix model was solved a long time ago \cite{brezin}. In
 this case the only unknown variable is the density of eigenvalues $\rho$. This
 model was solved using the SDe's in \cite{spenta}.

It is easy to show, using the properties of the projectors, that:
\beqn
u[\lambda]=\sum_\alpha e^{i\lambda_\alpha} T^\alpha\\
tr u[\lambda]=\sum_\alpha e^{i\lambda_\alpha}
\eeqn
That is, in $d=0$, $tr u[\lambda]$ decouples from $M$.
Replacing $tr u[\lambda]$ into the loop equation we get the equation that
 determines the eigenvalues:
\be
\sum_\alpha
 e^{i\lambda_\alpha}\left(2\sum_{\beta\neq\alpha}\frac{1}{m_\alpha-m_\beta}
-\Delta_\alpha\right)=0
\ee

Since $\lambda_\alpha$ is arbitrary, we obtain the known answer:
\be
\Delta_\alpha=2\sum_{\alpha\neq\beta}\frac{1}{m_\alpha-m_\beta}
\ee

\section{Discussion and Open Problems}

We have been able to derive the loop equations for the Hermitian matrix model
in
 arbitrary space-time dimension $d$ in terms of a functional $u[\lambda]$.
 Moreover, in leading order of large $N$, the loop equations form a closed set.
 All this was possible because we had chosen to write
the SDe's in the projector-eigenvalue basis instead of the $M$-basis, as
usually
 done in zero dimensions.

We believe this to be  an important step in the comprehension of the physics of
 this model. Although no solution of the equations is available at present, a
 numerical solution may be feasible. Also a variational
ansatz of the type employed in \cite{migdal1} for the QCD string may work.

{}From a conceptual point of view,  our loop equations may shed light into the
 geometrical structure of the model. The $d$-dimensional matrix model provides
a
 discretization  for  a string  in $d+1$ dimensions, but the exact geometrical
 mapping between the two descriptions has not been done yet. As in QCD
 \cite{migdal2}  may be the loop equations will help to bridge such a gap.

\section*{Acknowledgements}

The author acknowledges useful conversa\-tions with Luis Alvarez-\\Gaum\'e and
 Spenta Wadia. He also wants to thank Poul H. Damgaard for a reading of the
 manuscript.

This work has been partially financed by Fondecyt \#  1930566.

\setcounter{equation}{0}
\renewcommand{\thesection}{Appendix \Alph{section}.}
\renewcommand{\theequation}{A.\arabic{equation}}
\section*{Appendix}

In this Appendix we want to derive the Schwinger-Dyson BRST variation of the
 eigenvalues and projectors of the matrix $M$. Since the space-time dependence
 of the  involved variables is irrelevant to this discussion, we will not write
 it explicitly.

To begin with, we write the eigenvalue equation satisfied by the projectors:
\be
M T^\alpha=m_\alpha T^\alpha .
\ee
Computing the BRST variation of this identity, multiplying by $T^\beta$ from
the
 left and symmetrizing the resulting expression in $\alpha$ and $\beta$ we get:
\be
T^\alpha\psi T^\beta+T^\beta\psi T^\alpha=
(m_\alpha-m_\beta)(T^\beta\delta T^\alpha-T^\alpha\delta T^\beta)+
2\delta m_\alpha T^\alpha\delta_{\alpha\beta} .\label{ec}
\ee
{}From $\alpha=\beta$ we get
\be
T^\alpha\psi T^\alpha=\delta m_\alpha T^\alpha .
\ee
That is $\delta m_\alpha$ is the eigenvalue of $T^\alpha\psi$ of eigenvector
 $T^\alpha$. Computing the trace of the last equation we get:
\be
\delta m_\alpha= tr\psi T^\alpha .
\ee
Consider now the case $\alpha\neq\beta$ in (\ref{ec}). Dividing by
 $m_\alpha-m_\beta$
and summing over $\beta\neq\alpha$ we get the BRST variation of the projectors:
\be
\delta T^{\alpha} = \sum_{\beta \neq \alpha}\frac{T^{\alpha}
\psi T^{\beta} + T^{\beta}\psi T^{\alpha}}{ m_{\alpha}-m_{\beta}} .
\ee

As a check, we can easily show that these transformation rules for the
 projectors and eigenvalues of $M(x)$ are compatible with eqs.  (15)--(17) and
 that they reproduce  eq. (11) as well.
\end{document}